\documentclass[a4paper,aps,pra,preprintnumbers,amsmath,amssymb]{revtex4-2}
\usepackage{graphicx}
\usepackage{bm}
\usepackage{color}

\usepackage{amsmath}
\usepackage{amsfonts}
\usepackage{amssymb}
\usepackage{hyperref}
\usepackage{makeidx}
\usepackage{graphicx}
\usepackage{physics}
\usepackage{qcircuit}
\usepackage{natbib}

\begin{document}
	\title{Symmetric Bidirectional Quantum Teleportation using a Six-Qubit Cluster State  as a Quantum Channel}
	\author{Javid A Malik}
	\author{Muzaffar Qadir Lone}
	\author{Rayees A Mala} 
	\address{
		Department of Physics, University of Kashmir, Srinagar-190006, India	}

	\begin{abstract}
	Bidirectional quantum teleportation is a fundamental protocol for exchanging quantum information between two quantum nodes. All bidirectional quantum teleportation protocols till now have achieved a maximum efficiency of $40\%$. Here, we propose a new scheme for symmetric bidirectional quantum teleportation using a six-qubit cluster state as the quantum channel, for symmetric ($3\leftrightarrow3$) qubit bidirectional quantum teleportation of a special three-qubit entangled state. The novelty of our scheme lies in its generalization for ($N\leftrightarrow N$) qubit bidirectional quantum teleportation employing a $2N$-qubit cluster state. The efficiency of the proposed protocol is remarkably increased to $50\%$ which is highest till now. Interestingly, only GHZ-state measurements and four Toffoli-gate operations are necessary which is independent of the number $(N)$ of qubits to be teleported.
	\end{abstract}

	\maketitle

	\section{Introduction}
Quantum Teleportation (QT) first proposed by Bennett et al. \cite{bennett1}, is one of the most extraordinary applications of the quantum correlations i.e. entanglement \cite{Horode} present in a composite quantum system. It is a communication protocol that enables transmission of quantum information from one location to another, with the aid of preshared entanglement, local operations, and classical communications. 
Since its inception, many key developments in the QT have been reported \cite{werner2001,sasaki2014practical,yan2004scheme,man,bouwm,ma12,jin2010,zhang2006,pirandola,ren2017,yang2009,rigolin2005,li2019,agr2002,saha2012n,agrawal2006}.

Bidirectional quantum teleportation (BQT) \cite{zha2013bidirectional}  is an advanced version of QT and is defined as the bilateral QT protocol  between two quantum nodes in a quantum communication channel. This protocol finds its applications in cryptography  such as quantum secure direct communication (QSDC)\cite{wang2005multi}, quantum secret sharing (QSS), and cryptographic switch \cite{srinatha2014quantum}, etc.  BQT can be carried out with  or without  the aid of a controller. The first bidirectional quantum controlled teleportation (BQCT) protocol was introduced in \cite{zha2013bidirectional}, that employed a five-qubit cluster state for mutual transfer of an unknown single-qubit state between Alice and Bob with the help of controller Charlie. Several BQCT protocols have been proposed by utilizing multi-qubit cluster states as a quantum channel\cite{shukla2013bidirectional,chen2014bidirectional,duan2014bidirectional,sang2016bidirectional,zhou2019cyclic}. In\cite{duan2014bidirectional}  seven-qubit cluster state is used for asymmetric bidirectional quantum controlled teleportation where Alice sends an arbitrary single-qubit state to Bob and Bob sends a two-qubit state to Alice with the control of Charlie.

Several two-party BQT protocols have been proposed\cite{ty,zadeh2017bidirectional,javid2020,zhou2020quantum,zhou2019BQT,verma2020bidirectional,verma2020}. Shang et al. \cite{ty}, proposed an asymmetric bidirectional quantum teleportation protocol in absence of the controller. By utilizing a five-qubit cluster state as a channel, Alice teleports a two-qubit entangled state to Bob and Bob sends an arbitrary single-qubit state to Alice. Using the eight-qubit cluster state as the quantum channel Zadeh et al.\cite{zadeh2017bidirectional}  faithfully achieved the BQT of an arbitary two-qubit state. To achieve the BQT of multi-qubit states different cluster states were used. Chen et al.\cite{javid2020} utilised a four-qubit GHZ-state and two Bell states as a quantum channel to mutually teleport two arbitrary single-qubit states and an unknown three-qubit state between Alice and Bob. In\cite{zhou2019BQT}, a new BQT protocol was proposed using a six-qubit cluster state, however, it used non-local CNOT operation for its action, resulting in a decrease of intrinsic efficiency and an increase in operational complexity. This use of non-local CNOT operation was eliminated by Verma  \cite{verma2020bidirectional} by proposing a modified BQT protocol. In, order to employ N-qubit BQT Verma \cite{verma2020} proposed a protocol  with one of the  G-state\cite{rigolin2005} as a quantum channel. In this protocol, to achieve faithful transmission of N-qubits 2(2N-1)-CNOT gate operations along with two Hadamard gate operations are required. This increases the operational complexity without any improvement in intrinsic efficiency. It is therefore desirable to have a  new scheme that reduces the operational complexity by reducing the number of CNOT gate operations and increasing the intrinsic efficiency.

 Along this direction, to have greater efficiency and minimum complexity, we propose a new symmetric BQT protocol \ref{figure1} for mutual transmission of a special three-qubit entangled state using six-qubit cluster state as a quantum channnel. We also generalize our proposed protocol for implementation to N-qubit BQT. 
 The novelity of our scheme  as compared 
  to the previous  protocols\cite{ty,zadeh2017bidirectional,javid2020,zhou2020quantum,zhou2019BQT,verma2020bidirectional,verma2020}  is (i) only four Toffoli gate operations are required which is independent of the number (N) of qubits  to be teleported, (ii) operation complexity is greatly reduced, (iii) by introducing the auxiliary qubits the intrinsic efficiency is highest till now ($50\%$) achieved in any bidirectional quantum teleportation protocol.

This paper is organised as follows: In Section II, the proposed BQT protocol based on six-qubit cluster state is presented in detail. In Section III, we generalize our proposed protocol for N-qubit BQT. Finally, the we compare  our work with previous protocols and conclude in section IV.

\section{Protocol for Symmetric BQT ($3\leftrightarrow3$) Using Six Qubit Cluster State}

We propose a scheme for BQT that utilizes a six qubit cluster state as a quantum channel shared between Alice and Bob:
\begin{eqnarray}
	\ket{\phi}_{123456} 
	=\frac{1}{2}\Big(\ket{000000}+\ket{000111}  \nonumber +\ket{111000}+ \ket{111111}\Big)
\end{eqnarray}
where the indices $1-6$ refer to the ordering of qubits. 
Assume that Alice is in the possession of a special three-qubit entangled state, which is given by
\begin{eqnarray}
	\ket{\chi}_{a_{1}a_{2}a_{3}}=\large(\alpha\ket{000}+\beta\ket{111})_{a_{1}a_{2}a_{3}}
	\label{AQ}
\end{eqnarray}
with $ |\alpha|^{2}+|\beta|^{2}=1 $. At the same time, Bob also possesses a similar three qubit entangled state which can be expressed as
\begin{eqnarray}
	\ket{\chi}_{b_{1}b_{2}b_{3}}=\large(\gamma\ket{000}+\delta\ket{111})_{b_{1}b_{2}b_{3}}
	\label{BQ}
\end{eqnarray}
such that $ |\gamma|^{2}+|\delta|^{2}=1 $. Now, Alice and Bob wants to transmit the respective states to each other simultaneously, let the quantum channel given by Eq.(1) be distributed between Alice and Bob so that  the qubits 1, 4, 5 belongs to Alice and Bob holds the qubits 2, 3, 6. The proposed scheme for implementation of BQT ($3\leftrightarrow3$) is executed in the following manner:\\

\begin{figure}
	\includegraphics[width=10cm,height=6cm]{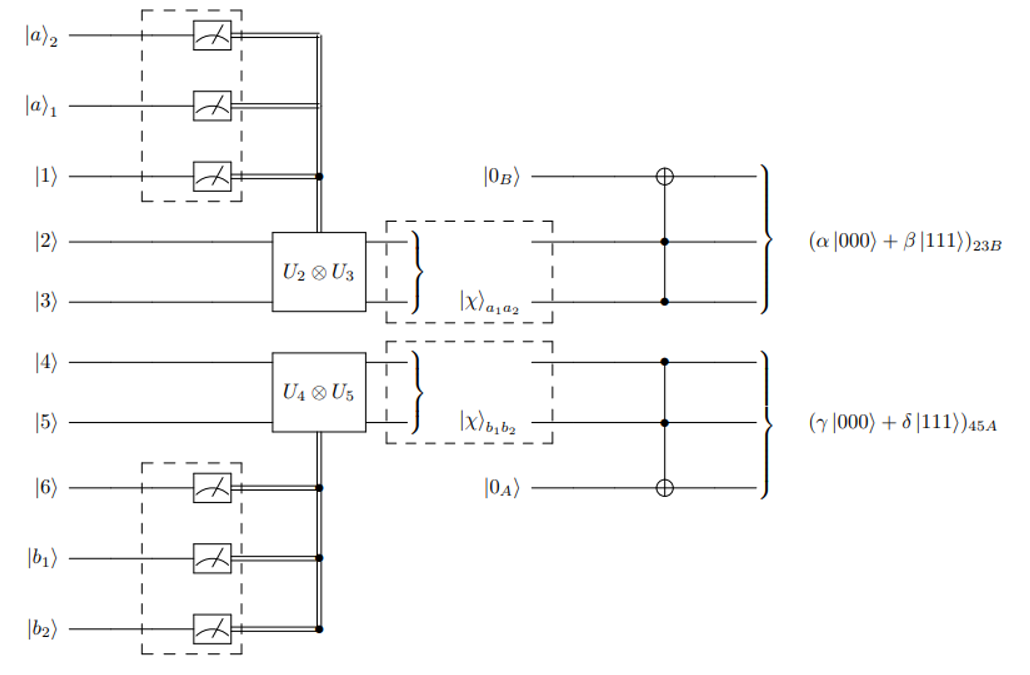}
	\caption{Schematic representation of the  Symmetric BQT proposed in this paper. Alice and Bob share a six qubit cluster state, and both parties send  three qubit states (equations \ref{AQ}-\ref{BQ}), bi-directionally. 
The upper dashed  boxes represent three-qubit GHZ-state measurement performed by Alice and the lower dashed  box represents Bob's three-qubit GHZ-state measurement. $ U_2\otimes U_3$ refers to recovery operations performed by Bob and $ U_4\otimes U_5$ by Alice as shown in Table-\ref{table2}. $\ket{\chi}_{a_{1}a_{2}} $ and $\ket{\chi}_{b_{1}b_{2}} $ are the two qubit entangled states recovered by Bob and Alice respectively. Using, only  two auxillary qubits $(\ket{0_A},\ket{0_B})$along with 	
 four Toffoli gate operations, the bidirectional teleportation is succesful with the utilization of minimum resources and maximum efficiency as compared to the previous proposed protocols. 
 $(\alpha |000\rangle + \beta |111\rangle )_{23B}$ and $(\gamma |000\rangle + \delta |111\rangle)_{45A}$ are the recovered three qubit states. }
	\label{figure1}
\end{figure}

{\bf \textit{Step-I:  Conversion of three-qubit entangled state to two-qubit entangled state and single-qubit state (Toffoli-gate operations):}}
Toffoli-gate (also known as CCNOT-gate) is a three-qubit gate with two controls and one target. It performs an X on the target only if both controls are in the state $|1\rangle$.

\begin{figure}[h]
	\includegraphics[width=8cm,height=4cm]{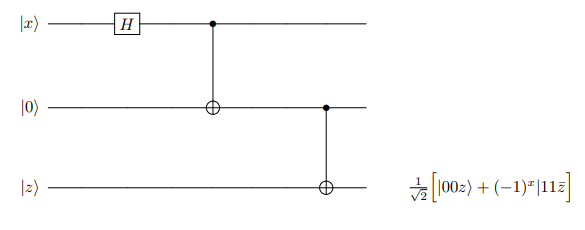}
	\caption{Quantum circuit to obtain the GHZ-states used for measurement by Alice and Bob given in equations \ref{g1} to \ref{g4}.
}
	\label{figure5}
\end{figure}

\begin{itemize}
	\item  Alice applies Toffoli-gate  with $a_{1}a_{2}$ as control and $a_{3}$ as target such that 
	$\ket{\chi}_{a_{1}a_{2}a_{3}}\longrightarrow\ket{\chi}_{a_{1}a_{2}}\otimes\ket{0}_{a_{3}}$

	with
	\begin{equation}
		\ket{\chi}_{a_{1}a_{2}}=\large(\alpha\ket{00}+\beta\ket{11})_{a_{1}a_{2}}
	\end{equation} 
	\item Next, Bob applies  Toffoli-gate with $b_1b_{2}$ as control and $b_{3}$ as target such that 
	$\ket{\chi}_{b_{1}b_{2}b_{3}}\longrightarrow\ket{\chi}_{b_{1}b_{2}}\otimes\ket{0}_{b_{3}}$
	
	with
	\begin{eqnarray}
		\ket{\chi}_{b_{1}b_{2}}=\large(\gamma\ket{00}+\delta\ket{11})_{b_{1}b_{2}}.
	\end{eqnarray} 
\end{itemize}  
After this step the joint state of the qubits $a_{1}a_{2}b_{1}b_{2}123456$ becomes   
\begin{eqnarray}
	\ket{\psi}_ {a_{1}a_{2}b_{1}b_{2}}^{123456 } 
	&=&\ket{\chi}_{a_{1}a_{2}}\otimes\ket{\chi}_{b_{1}b_{2}} \otimes\ket{\phi}_{123456} \nonumber \\
	&=&\frac{1}{2}\big(\alpha\ket{00}+\beta\ket{11})_{a_{1}a_{2}}\otimes\large(\gamma\ket{00}+\delta\ket{11})_{b_{1}b_{2}}  \otimes 
	\large(\ket{000000}+\ket{000111}+\ket{111000}+\ket{111111})_{123456}
\end{eqnarray} 
The joint state $\ket{\psi}_ {a_{1}a_{2}b_{1}b_{2}}^{123456 } $ takes the form
\begin{eqnarray}
	\ket{\psi}_ {a_{1}a_{2}b_{1}b_{2}}^{123456 }&=&\frac{1}{4}\Big[\ket{\eta_{1}^{\pm}}\ket{\zeta_{1}^{\pm}} \ket{\psi_{1}^{\pm}}+\ket{\eta_{1}^{\pm}}\ket{\zeta_{1}^{\mp}} \ket{\psi_{2}^{\pm}} + \ket{\eta_{1}^{\pm}}\ket{\zeta_{2}^{\pm}} \ket{\psi_{3}^{\pm}}+\ket{\eta_{1}^{\pm}}\ket{\zeta_{2}^{\mp}} \ket{\psi_{4}^{\pm}} \nonumber \\
	&& + 
	\ket{\eta_{2}^{\pm}}\ket{\zeta_{1}^{\pm}} \ket{\psi_{5}^{\pm}}+\ket{\eta_{2}^{\pm}}\ket{\zeta_{1}^{\mp}} \ket{\psi_{6}^{\pm}} +
	\ket{\eta_{2}^{\pm}}\ket{\zeta_{2}^{\pm}} \ket{\psi_{7}^{\pm}}+\ket{\eta_{2}^{\pm}}\ket{\zeta_{2}^{\mp}} \ket{\psi_{8}^{\pm}}\Big]
\end{eqnarray}
where $\ket{\eta_{i}^{\pm}} $ and $\ket{\zeta_{i}^{\pm}}$ are GHZ states \ref{figure5} given by 

\begin{eqnarray}
	\ket{\eta_{1}^{\pm}}_{a_{1}a_{2}1}&=\frac{1}{\sqrt{2}}\Big(\ket{000}\pm\ket{111}\Big)\label{g1}&\\  \ket{\eta_{2}^{\pm}}_{a_{1}a_{2}1}&=\frac{1}{\sqrt{2}}\Big(\ket{001}\pm\ket{110}\Big)\label{g2}\\
	\ket{\zeta_{1}^{\pm}}_{b_{1}b_{2}6}&=\frac{1}{\sqrt{2}}\Big(\ket{000}\pm\ket{111}\Big)\label{g3}\\
	\ket{\zeta_{2}^{\pm}}_{b_{1}b_{2}6}&=\frac{1}{\sqrt{2}}\Big(\ket{001}\pm\ket{110}\Big)\label{g4}
\end{eqnarray}

and $\ket{\psi_{i}}^{\pm}$ are the states of the qubits 2, 3, 4 and 5 shown in Table-\ref{table1}. 

{ \bf \textit{Step-II: Measurements and classical communications}}
\begin{itemize}
	\item 
	Alice performs a GHZ-state measurement(GSM) on her qubits ($a_{1}a_{2}1$) in the measurement basis made up of the states $ \ket{\eta_{i}}^{\pm},i=1,2.$
\end{itemize}
\begin{itemize}
	\item 
	Bob also performs a GHZ-state measurement(GSM) on his qubits ($b_{1}b_{2}6$) in measurement basis consisting of the states $ \ket{\zeta_{i}}^{\pm},i=1,2.$ 
\end{itemize}

Afterwards Alice and Bob convey their measurement outcomes to each other by classical communication, and corresponding to each measurement outcome is a collapsed state of the qubits 2, 3, 4 and 5. In, order to recover the target states Alice and Bob apply the local operations on their qubits. The collapsed states of particles 2, 3, 4 and 5 after GHZ-state measurements(GSM) are given in Table-I.

\begin{table}[t]
	\begin{tabular}{|c|c|}
		\hline
		$\ket{\psi_{1}^{\pm}}$  & $ = \big(\alpha\ket{00} \pm\beta\ket{11}\big)_{23}\otimes\big(\gamma\ket{00} \pm\delta\ket{11}\big)_{45}$ \\ \hline   
		$\ket{\psi_{2}^{\pm}}$  & $ = \big(\alpha\ket{00} \pm\beta\ket{11}\big)_{23}\otimes\big(\gamma\ket{00} \mp\delta\ket{11}\big)_{45}$ \\   \hline  
		$\ket{\psi_{3}^{\pm}}$  & $ = \big(\alpha\ket{00} \pm\beta\ket{11}\big)_{23}\otimes\big(\gamma\ket{11} \pm\delta\ket{00}\big)_{45}$ \\      \hline
		$\ket{\psi_{4}^{\pm}}$  & $ = \big(\alpha\ket{00} \pm\beta\ket{11}\big)_{23}\otimes\big(\gamma\ket{11} \mp\delta\ket{00}\big)_{45}$ \\ \hline
		$\ket{\psi_{5}^{\pm}}$  & $ = \big(\alpha\ket{11} \pm\beta\ket{00}\big)_{23}\otimes\big(\gamma\ket{00} \pm\delta\ket{11}\big)_{45}$ \\  \hline
		$\ket{\psi_{6}^{\pm}}$  & $ = \big(\alpha\ket{11} \pm\beta\ket{00}\big)_{23}\otimes\big(\gamma\ket{00} \pm\delta\ket{11}\big)_{45}$ \\  \hline     
		$\ket{\psi_{7}^{\pm}}$  & $ = \big(\alpha\ket{11} \pm\beta\ket{00}\big)_{23}\otimes\big(\gamma\ket{11} \pm\delta\ket{00}\big)_{45}$ \\  \hline      
		$\ket{\psi_{8}^{\pm}}$  & $ = \big(\alpha\ket{11} \pm\beta\ket{00}\big)_{23}\otimes\big(\gamma\ket{11} \mp\delta\ket{00}\big)_{45}$ \\
		\hline    
	\end{tabular}
	\caption{ The collapsed states of the qubits 2, 3, 4 and 5 after GHZ-state measurements.}
	\label{table1}
\end{table}

{\bf \textit{Step-III: Local Operations}}
\begin{itemize}
	\item Alice applies appropriate unitary operation($U_4 \otimes U_5$) to recover the two qubit state depending upon classical information received from Bob's measurement.
	\item  Bob also applies appropriate unitary operation($U_2 \otimes U_3$) to recover the two qubit state depending upon Alice's measurement outcome.
\end{itemize}
The GHZ-state measurement outcomes of Alice and Bob and corresponding unitary operations are given in Table-\ref{table2}.
In order to recover the three qubit target entangled states, once two qubit entangled states are recovered by Alice and Bob, we introduce two auxiliary qubits $\ket{0}_{A}$ and $\ket{0}_{B}$.

\begin{table}[htbp]
	\begin{tabular}{|p{1cm}|p{1cm}|p{1.3cm}|p{1.5cm}|p{1.5cm}|}
		\hline
		Alice's MR &Bob's MR&Collapsed state&Alice,s UO ($U_4 \otimes U_5$)&Bob's UO ($U_2 \otimes U_3$)  \\
		\hline
		$ \ket{\eta_{1}^{+}} $ &  $ \ket{\zeta_{1}^{+}} $& $ \ket{\psi_{1}^{+}} $&$\sigma_{0}\otimes\sigma_{0}$&$\sigma_{0}\otimes\sigma_{0}$\\ \hline
		$ \ket{\eta_{1}^{-}} $ &  $ \ket{\zeta_{1}^{-}} $& $ \ket{\psi_{1}^{-}} $&$\sigma_{0}\otimes\sigma_{3}$&$\sigma_{0}\otimes\sigma_{3}$\\ \hline
		$ \ket{\eta_{1}^{+}} $ &  $ \ket{\zeta_{1}^{-}} $& $ \ket{\psi_{2}^{+}} $&$\sigma_{0}\otimes\sigma_{3}$&$\sigma_{0}\otimes\sigma_{0}$\\ \hline
		$ \ket{\eta_{1}^{-}} $ &  $ \ket{\zeta_{1}^{+}} $& $ \ket{\psi_{2}^{-}} $&$\sigma_{0}\otimes\sigma_{0}$&$\sigma_{3}\otimes\sigma_{0}$\\ \hline
		$ \ket{\eta_{1}^{+}} $ &  $ \ket{\zeta_{2}^{+}} $& $ \ket{\psi_{3}^{+}} $&$\sigma_{1}\otimes\sigma_{1}$&$\sigma_{0}\otimes\sigma_{0}$\\    \hline
		$ \ket{\eta_{1}^{-}} $ &  $ \ket{\zeta_{2}^{-}} $& $ \ket{\psi_{3}^{-}} $&$\sigma_{1}\otimes i\sigma_{2}$&$\sigma_{3}\otimes\sigma_{0}$\\ \hline
		$ \ket{\eta_{1}^{+}} $ &  $ \ket{\zeta_{2}^{-}} $& $ \ket{\psi_{4}^{+}} $&$ i\sigma_{2}\otimes\sigma_{1}$&$\sigma_{0}\otimes\sigma_{0}$\\ \hline
		$ \ket{\eta_{1}^{-}} $ &  $ \ket{\zeta_{2}^{+}} $& $ \ket{\psi_{4}^{-}} $&$i\sigma_{2}\otimes i \sigma_{2}$&$\sigma_{0}\otimes\sigma_{3}$\\ \hline
		$ \ket{\eta_{2}^{+}} $ &  $ \ket{\zeta_{1}^{+}} $& $ \ket{\psi_{5}^{+}} $&$\sigma_{0}\otimes\sigma_{0}$&$\sigma_{1}\otimes\sigma_{1}$\\ \hline
		$ \ket{\eta_{2}^{-}} $ &  $ \ket{\zeta_{1}^{-}} $& $ \ket{\psi_{5}^{-}} $&$\sigma_{0}\otimes\sigma_{3}$&$i\sigma_{2}\otimes\sigma_{1}$\\ \hline
		$ \ket{\eta_{2}^{+}} $ &  $ \ket{\zeta_{1}^{-}} $& $ \ket{\psi_{6}^{+}} $&$\sigma_{3}\otimes\sigma_{0}$&$ i\sigma_{2}\otimes i\sigma_{2}$\\ \hline
		$ \ket{\eta_{2}^{-}} $ &  $ \ket{\zeta_{1}^{+}} $& $ \ket{\psi_{6}^{-}} $&$\sigma_{0}\otimes\sigma_{0}$&$\sigma_{1}\otimes i\sigma_{2}$\\ \hline
		$ \ket{\eta_{2}^{+}} $ &  $ \ket{\zeta_{1}^{+}} $& $ \ket{\psi_{5}^{+}} $&$\sigma_{0}\otimes\sigma_{0}$&$\sigma_{1}\otimes\sigma_{1}$\\ \hline
		$ \ket{\eta_{2}^{+}} $ &  $ \ket{\zeta_{2}^{+}} $& $ \ket{\psi_{7}^{+}} $&$\sigma_{1}\otimes\sigma_{1}$&$\sigma_{1}\otimes\sigma_{1}$\\ \hline
		$ \ket{\eta_{2}^{-}} $ &  $ \ket{\zeta_{2}^{-}} $& $ \ket{\psi_{7}^{-}} $&$\sigma_{1}\otimes i\sigma_{2}$&$ i\sigma_{2}\otimes\sigma_{1}$\\ \hline
		$ \ket{\eta_{2}^{+}} $ &  $ \ket{\zeta_{2}^{-}} $& $ \ket{\psi_{8}^{+}} $&$i\sigma_{2}\otimes\sigma_{1}$&$i\sigma_{2}\otimes i\sigma_{2}$\\\hline
		$ \ket{\eta_{2}^{-}} $ &  $ \ket{\zeta_{2}^{+}} $& $ \ket{\psi_{8}^{-}} $&$i\sigma_{2}\otimes i\sigma_{2}$&$\sigma_{1}\otimes i\sigma_{2}$\\
		\hline
	\end{tabular}
	\caption{Measurement outcomes, corresponding unitary operations and collapsed states.}
	\label{table2}
\end{table}

{\bf \textit{Step-IV: Toffoli-gate operations}}

\begin{itemize}
	\item Alice applies Toffoli gate with qubits 4, 5 as control and $\ket{0}_{A}$ as target qubit to recover the target three qubit entangled state \big($\gamma\ket{000}+\delta\ket{111}\big)_{45A} $ shown in Figure- \ref{figure1}   
	\item Bob also applies Toffoli gate with qubits 2, 3 as control and $\ket{0}_{B}$ as target qubits to recover the desired state  \big($\alpha\ket{000}+\beta\ket{111}\big)_{23B} $ shown in Figure-\ref{figure1}  
\end{itemize}

Thus, symmetric ($3\leftrightarrow3$) BQT protocol is faithfully and deterministically observed. 

\section{Generalization to $N$-Qubits}
In this section our proposed scheme for symmetric BQT can be generalised for ($ N\leftrightarrow N$) qubit BQT using the 2N-qubit cluster state as a quantum channel. In, our scheme no additional quantum resources are required which makes it a novel scheme for BQT. Let Alice and Bob are respectively in a possession of the following $N$-qubit entangled states

\begin{eqnarray}
	\ket{\chi}_{a_{1}a_{2}\dots a_{N}}&=&(\alpha\ket{00\dots0}+\beta\ket{11\dots1})_{a_{1}a_{2}\dots a_{N}} \\
	\ket{\chi}_{b_{1}b_{2}\dots b_{N}}&=&(\gamma\ket{00\dots0}+\delta\ket{11\dots1})_{b_{1}b_{2}\dots b_{N}}
\end{eqnarray} 
By utilising the $2N$-qubit cluster state and employing our scheme Alice and Bob can faithfully achieve mutual transmission of the above $N$-qubit states. The $2N$-qubit quantum channel has the form
\begin{eqnarray}
	\ket{\phi}_{12\dots2N}=\frac{1}{{2}}\bigg(\ket{x_{1}}+\ket{x_{2}}+\ket{x_{3}}+\ket{x_{4}}\bigg)_{12\dots2N}
\end{eqnarray}
where 
\begin{eqnarray}
	&&	\ket{x_{1}}=\otimes_{i=1}^{N}\ket{0}\otimes_{i=N}^{2N}\ket{0}, ~~~~~~\ket{x_{2}}=\otimes_{i=1}^{N}\ket{1}\otimes_{i=N}^{2N}\ket{1} \\
	&&	\ket{x_{3}}=\otimes_{i=1}^{N}\ket{0}\otimes_{i=N}^{2N}\ket{1}, ~~~~~~\ket{x_{4}}=\otimes_{i=1}^{N}\ket{1}\otimes_{i=N}^{2N}\ket{0}
\end{eqnarray}
Let Alice has qubits $ 1, N+1 ,\dots, 2N-1 $ and Bob has the qubits $2, 3, \dots, N, 2N.$ The steps involved in ($ N\leftrightarrow N$)qubit symmetric BQT protocol are same as that of the ($3\leftrightarrow3$) qubit BQT protocol. The implementation of our scheme for general case is described in following steps.

{\textit{ First step}}: Alice applies a Toffoli gate operation with ($a_{N-2} ,a_{N-1}$) as control qubits and $ a_N$ as target qubit to convert the N-qubit state to be teleported into an $N-1$ qubit entangled state and a single qubit. Bob also applies Toffoli gate with ($b_{N-2} ,b_{N-1}$) as control qubits and $ b_N$ as target qubit. After this step the joint state of the (4N-2) qubits ($a_1a_2\dots a_{N-1}b_1b_2\dots b_{N-1}12\dots 2N $) becomes     
\begin{eqnarray}
	\ket{\psi}=\ket{\chi}_{a_{1}a_{2}\dots a_{N-1}}\otimes\ket{\chi}_{b_{1}b_{2}\dots b_{N-1}}\otimes\ket{\phi}_{12\dots2N}
\end{eqnarray}

{\textit{ Second step}}: Alice performs $N$-qubit GHZ-state measurement (NGSM) on qubits ($a_{1}a_{2}\dots a_{N-1},1 )$ and convey the measurement outcomes to Bob.  Bob also performs NGSM on qubits ($b_{1}b_{2}\dots b_{N-1},2N )$ and convey the outcomes to Alice. 

{\textit{Third step}}: Alice and Bob apply the unitary operations on their qubits ($N+1,\dots, 2N-1 $) and( $2,3,\dots,N$) respectively depending upon their measurement outcomes enabling them to recover the corresponding $(N-1)$ qubit entangled states.

{\textit{Fourth step}}:In order to recover the target $N$-qubit entangled states, two auxiliary qubits $\ket{0}_{A}$ and  $\ket{0}_{B}$ are introduced as in section-II. Alice applies the Toffoli gate with qubits $ (2N-2, 2N-1 ) $as control and $\ket{0}_{A}$ as target qubit. Bob also applies Toffoli gate with  $(N-1, N )$ as control qubits and  $\ket{0}_{B}$ as target qubit.

In, this manner our scheme for symmetric BQT can be generalised for ( $N\leftrightarrow N $) qubit BQT successfully. Only four Toffoli-gate operations are required for its implementation, which is interestingly same for general case i.e, independent of number of qubits to be teleported.

\begin{table}[h]
	\begin{tabular}{|p{1.2cm}|p{.5cm}|p{2cm}|p{1cm}|p{2mm}|c|c|}
		\hline    
		Protocol&$q_s$&NO&$b_t$&$q_a$&BQT&$\eta $\\
		\hline 
		\cite{ty}  & $5$&$2BSM$&$4$&$0$&$2\leftrightarrow1$&$33.3$\\
		\hline
		\cite{javid2020}  & $8$&$4SM,4QM$&$8$&$0$&$2\leftrightarrow3$&$31.25$\\
		\hline
		\cite{zhou2020quantum} & $6$&$2GSM$&$4$&$0$&$2\leftrightarrow2$&$40$\\
		\hline
		\cite{zhou2019BQT} & $7$&1GSM,1BSM, 1SM &$7$&$0$&$3\leftrightarrow1$&$28.5$\\
		\hline
		\cite{verma2020bidirectional} & $6$&1GSM, 1BSM&$4$&$0$&$3\leftrightarrow1$&$40$\\
		\hline
		\cite{verma2020} & $4$&$2MCB$&$4$&$2$&$2\leftrightarrow2$&$40$\\
		\hline
		${\rm Our} $ & $6$&$2GSM$&$4$&$2$&$3\leftrightarrow3$&$50$\\
		\hline
		${\rm Our} $ & $2N$&$2GSM$&$2N-2$&$2$&$N\leftrightarrow N$&$50$\\
		\hline
	\end{tabular}   
	\caption{Comparison Among  BQT protocols. Here NO means number of operations required.} 
	\label{table3}
\end{table}

\section{Comparison and Conclusions}
To show the novelty of our BQT scheme, let us compare our protocol with the previous protocols as  shown in Table-\ref{table3}. 
The comparison is based on following aspects: the quantum  bits transmitted $q_t$, the quantum resources consumed ( number of qubits of the quantum channel $q_s$), the classical resources transmitted $b_t$, the number of auxiliary qubits $q_a$, the required operations (NO) and the intrinsic efficiency ($\eta$). The  intrinsic efficiency ($\eta$)  is defined as 
\begin{eqnarray}
\eta=\frac{q_t}{q_{s}+q_{a}+b_{t}} .
\end{eqnarray} 
It is clear that as the number of qubits to be teleported increases both the number of quantum bits of the channel and the classical resource consumption increases which results in the decrease of intrinsic efficiency. However, in our scheme by introducing the auxiliary qubits no additional quantum resources are required for implementation to the general case and the efficiency is also remarkably increased.

In conclusion, we have proposed a novel scheme for symmetric BQT via six-qubit cluster state as a quantum channel. Furthermore, we generalized our scheme to implement ($N\leftrightarrow N$) qubit BQT using the $2N$-qubit cluster state as the quantum channel. The proposed protocol is based on GHZ-state measurements (GSM), local unitary operations and Toffoli-gate operations. In, our scheme Alice and Bob are interested in mutual transmission of three-qubit entangled states without a controller. We employ Toffoli-gate operations to transform a three qubit entangled state into a two qubit entangled state and a single qubit state. At the end of the protocol we again perform Toffoli gate to restore the target three-qubit states. The previous protocol for $N$-qubit BQT \cite{verma2020}, employs $2(2N-1)$ CNOT-gate operations, Hadamard gate operations for its implementation. In our, scheme only Four Toffoli gate operations are necessary for $N$-qubit BQT thereby reducing the operational complexity and quantum resources involved in the BQT scheme. From Table-\ref{table3}, its clear that our scheme has the prominent advantage of more qubits transmitted and highest intrinsic efficiency.
 The extension of  our proposed protocol for BQT in noisy channels and influence of auxillary qubits on efficiency will be presented elsewhere \cite{Javid2022}.

\end{document}